\documentstyle[aps,prb,preprint]{revtex}
\title{Multiphonon decay of strong mode in quantum lattice}
\author{V.Hizhnyakov}
\address{Institute of Theoretical Physics, University of Tartu,
T\"ahe 4, EE2400 Tartu, Estonia \\
Institute of Physics, Riia 142, EE2400 Tartu, Estonia\\
e-mail: hizh@fi.tartu.ee}
\begin{document}
\maketitle
\begin{abstract}
A nonperturbative theory of multiphonon anharmonic decay of strongly
excited local mode is developed whereby the mode is considered classically
and phonons, quantum mechanically. The decay rate of the mode is expressed
via the negative frequency parts of the phonon pair correlation functions.
In the case of two-phonon decay the later satisfy the linear integral
equations while in the case of two- and more-phonon decay they satisfy the
nonlinear integral equations. As a result, the processes
mentioned differently depend on the mode amplitude $A$: two-phonon processes
smoothly deminish if $A \rightarrow \infty$ while
three- and more-phonon processes are fully switched-off at
large amplitudes and they abruptly switch-on if the amplitude
approaches the critical value. At that the decay rate gets rather
high value (of the order of the mode quantum per period). The final
stage of the relaxation is well described by the perturbation theory.
\end{abstract}

\section{Introduction}
The problem of the time evolution of strong vibrational excitations in
crystals nowadays attracts a remarkable attention. One of the essential
results in this field is the demonstration of the existence of localized
modes (ILM, intrinsic local mode or SLS, self-localized soliton
\cite{dolgov,sivtak,page,burlakov,zavt,kiselev,sanpage,wagner})
in perfect crystals. The
results are obtained by applying equations of classical nonlinear
dynamics.
The central point concerning the ILMs is their stability. In the
classical limit these modes are stable if all their harmonics
are out of resonance with the phonon spectrum. This condition is fulfilled
e.g. for the local modes with high frequency (the basic frequency of the
mode is above the phonon band). On the contrary, the quantum mechanics
allows the distinction of the high frequency quanta (due to the processes of
creation of two or more phonons). Therefore, the question of the
stability of the modes mentioned should be considered on the basis of
the quantum theory. However, the problem mentioned can not be solved by
applying standard quantum perturbation theory, commonly used for the
description of quantum transitions since one cannot assume that the
perturbation parameter is small (note that one of the essential conditions
of existence of ILMs in a nonlinear lattice is a large amplitude of
vibration).

Recently we proposed a nonperturbative theory of the two-phonon anharmonic
decay of strong local modes associated with lattice defects \cite{hizh1}
and an analogous theory of the decay of the ILMs in the perfect
monatomic chain
\cite{hizh2}. We have found that the decay law is highly nonexponential:
at definite "critical" amplitudes (typically of the order of
0.2 - 0.4 \AA) the relaxation jumps take place being
accompanied by generation of phonon bursts. The process considered in
\cite{hizh1,hizh2} takes into account cubic anharmonicity.
The interaction term which is
responsible for the process is $\sim \sum_{i_1i_2}
W^{(3)}_{i_1 i_2} Q(t) \hat{x}_{i_1}
\hat{x}_{i_2}$, where $W^{(3)}_{i_1 i_2}$ are the parameters of the cubic
anharmonicity, $Q(t)$ is the time-dependent classical coordinate
of the strong local mode, $\hat{x}_{i_1}$ and  $\hat{x}_{i_2}$ are
the coordinate operators of phonons. Such interaction leads to the quadratic
(with respect to phonon operators $\hat{a}_i$, $\hat{a}_i^+$)
time-dependent phonon Hamiltonian. The latter can be diagonalized exactly
by means of the linear transformation of phonon operators
\begin{equation}
\hat{b}_j(t) = \sum_i(\mu_{ij}\hat{a}_i + \nu_{ij}\hat{a}_i^+);
\end{equation}
the terms $\sim\hat{a}_i^+$ result from fast (oscillatory) change in
time of the phonon Hamiltonian \cite{hizh1}. Due to these terms the
initial zero-point state $|0\rangle$ is not the zeroth state of
the transformed destruction operator $\hat{b}_j$, i.e.
phonons are created in the lattice.
This transformation allows one to find
energy of generated phonons and, thus the energy loss of the mode for
any cubic anharmonicity and at any time moment.

Let us mention that an analogous transformation of the field operators
takes place
in the field of a gravitationally collapsing star (black hole),
causing thermal radiation (so-called quantum emission of black holes)
\cite{hawking,birel,grib}.
Note also an analogy of emission of phonon pairs by a strong mode
with superconducting transition:
in both cases mean values of destruction (and creation) operators
(of phonons or electrons (holes), respectively) equal to zero but
mean values of products of two destruction operators (as well as mean values
of products of two creation operators) differ from zero.

The described process of phonon generation by a strong mode
is allowed if $\omega_l < 2 \omega_M$, where
$\omega_l$ is the frequency of the local mode,
$\omega_M$ is the top band frequency. If $\omega_l > 2\omega_M$, then one
needs to account for three- (or more) phonon processes. These processes
arise from the interaction terms
$\sim \sum_k W^{(n+k)}_{i_{1}i_{2}...i_{n-1}} Q^{k}(t) \hat{x}_{i_1}
\hat{x}_{i_2} \ldots \hat{x}_{i_{n-1}}$ with $n \geq 4$. The phonon
Hamiltonian
with such terms cannot be diagonalized. Therefore, the method of \cite{hizh1}
is not directly applicable for $n \geq 4$. Below we present another method
which is based on the asymptotical behaviour of a
pair correlation function and which is applicable for arbitrary $n$.

\section{General Formulas}
We take into account that according to the Coleman theorem \cite{coleman}
the vacuum (zero-point) state of bosons in the classical field is not
invariant with respect to the transformation which changes the classical field.
This means that destruction operators are different for different values of
the classical field. Therefore, if one describes the effect of a classical
mode, anharmonically interacting with phonons in the quasi-harmonic
(Hartree-Fock, or mean-field) approximation
\begin{equation}
\hat{H}_{ph} (t) = \sum_j \hbar \omega_j (\hat{b}_j^+ (t) \hat{b}_j (t) + 1/2)
\end{equation}
then the destruction operators linearly change with the mode amplitude. For
slow relaxation (as compared to the mode frequency $\omega_l$) the
transformation is diagonal in the phonon space:$$
\hat{b}_i (t) \simeq \mu_i (t) \hat{a}_i + \nu_i (t) \hat{a}_i^+$$
($\mu_{ij} \simeq \mu_i\delta_{ij}, \nu_{ij} \simeq \nu_i\delta_{ij}$; see
(1)). Here $\hat{a}_i$ and $\hat{a}^+_i$ are phonon operators before
excitation of the local mode (which is supposed to take place at time $t=0$);
$|\mu_i^2| - |\nu_i^2| = 1$; the constant term on the right hand side of (2)
is considered to be small and neglected. In this approximation
all multiphonon decay processes are described in the same way: the relaxation
rate of the mode
\begin{equation}
\frac{dE_l(t)}{dt} = - \frac{dE_{ph}(t)}{dt} = -\sum_i
\hbar \omega_i \frac{d|\nu_i(t)|^2}{dt}
\end{equation}

is determined by the negative frequency terms in the large time asymptotics
of the two-time correlation function
\begin{equation}
<0|\hat{x}_i (t + \tau) \hat{x}_i (t) |0> \simeq
\frac{\hbar}{2\omega_i}[|\mu_i(t)|^2 e^{-i\omega_i \tau} +
|\nu_i (t)|^2 e^{i\omega_i \tau} ], \;\; t \gg \tau \sim
 \omega_l^{-1}
 \end{equation}
(here fast oscillating terms $\sim \exp{(\pm i\omega_i (2t+\tau))}$
are averaged out).

$|\nu_i (t)|^2$ can be found as follows. 
Presenting coordinate operators
by means of initial creation and destruction operators
$\hat{x}_i(t) = \sqrt{\hbar/2\omega_i}(g_i(t)\hat{a}_i +
g_i^* (t) \hat{a}_i^+)$ one gets
\begin{equation}
\langle 0|\hat{x}_i (t+\tau) \hat{x}_i (t) |0 \rangle = (\hbar/2\omega_i)
g_i(t+\tau) g_i^*(t) .
\end{equation}
The time dependence of the coordinate operators can be determined from
the Lagrangian equations of motion.
Inserting these equations to the correlation functions (5) and
expressing multiphonon correlation functions via the Wick-type
products of pair correlation functions one
gets linear (for cubic anharmonicity) or nonlinear (for quartic and
higher order anharmonicity) equations for the pair correlation
functions. It is
straightforward to show that for the multiphonon decay
one obtains the same equations if to calculate $g_i(t)$ from
classical equations of motion for dimensionless normal displacements with
"nonclassical" initial conditions $g_i(t) = \exp{(-i\omega_i t + \phi_i)},
t\leq 0$
and then to average the correlation functions over the random phases $\phi_i$
(the later averaging is denoted below as $\langle...\rangle$).

We demonstrate the method for anharmonic interactions in  the
pair potential approximation with account of the central forces only.
For the case of simplicity we suppose
that the strong odd mode is fully localized on the site $l=0$.
Taking $Q(t) = A_t \cos{\omega_lt}$ one gets the following
interaction Hamiltonian
\begin{equation}
\hat{H}^{(n+1)}_{anh} \simeq \frac{2}{n} A_t\cos{\omega_lt} \sum_m V_m^{(n)}
\hat{q}^n_m, \, \,n \geq 2 ,
\end{equation}
where $V_m^{(n)} \equiv V_{mt}^{(n)} = - \sum_k W_{0m}^{(n+k+1)}A_t^k/(2(n-1)!)$
and $\omega_l \equiv \omega_{lt}$ depend on the mode amplitude,
$q_m(t) = \sum_i e_{mi}
\hat{x}_i(t)$.
Lagrangian equations for
$g_i (t)$ are the following:$$
\ddot{g}_i(t) + \omega_i^2 g_i(t) + 2A_t\cos{\omega_lt} \sum_m \omega_i \bar{e}_{im} V_m^{(n)}
g_m^{n-1}(t) = 0$$
($\bar{e}_{mi} = e_{mi}/\omega_i^{1/2}$).
Corresponding integral equations read
\begin{equation}
g_i(t) = e^{-i\omega_i t + \phi_i} + 2A_t\sum_m \bar{e}_{mi}V_m^{(n)}
\int_0^t dt_1 \sin(\omega_i(t-t_1))g_m^{n-1}(t_1)\cos{\omega_lt_1}
\end{equation}
where
$g_m(t) = \sum_i\bar{e}_{im} g_i(t)$ satisfy the analogous equations
\begin{equation}
g_m(t) = f_m(t) + 2A_t\sum_{m'} \int_0^t dt_1 G_{mm'} (t-t_1)
V_{m'}^{(n)}
g^{n-1}_{m'}(t_1)\cos{\omega_lt_1} ,
\end{equation}
$G_{mm'}(t) = \sum_i \bar{e}_{im} \bar{e}_{im'} \sin{\omega_i t}$ is the
dynamical Green function \cite{maradudin,econ}, $f_{m}(t) = \sum_i\bar{e}_
{im}exp(-i\omega_{i}t+\phi_i)$.

Substituting (7) to (5) and taking into account (3) and (4) one gets
for $t \gg \omega_{l}^{-1}$
\begin{equation}
\frac{dE_{ph}(t)}{dt}  = \frac{\hbar A^2_t(n-1)}{2\pi}
\int_0^{\omega_M}d\omega \omega \mbox{Sp}[ImG(\omega_l - \omega ) V^{(n)}
D^{(n-1)}(t;\omega) V^{(3)}],
\end{equation}
where
\begin{equation}
D_{mm'}^{(n-1)}(t;\omega) =
\int_{-\infty}^{\infty}d\tau e^{i\omega\tau}D^{n-1}_{mm'}(t;\tau),
\end{equation}
is the positive frequency $(\omega > 0 )$ part of the $n-1$ - phonon
spectrum;
\begin{equation}
D_{mm'}(t;\tau) = \langle g_m(t+\tau)g^*_{m'}(t) \rangle ,
\end{equation}
$$
G_{mm'}(\omega) = \int_{0}^{\infty} e^{-i\omega t} G_{mm'}(t)dt$$
is the phonon Green function in frequency representation. In (9) it is
accounted that the main contribution to the relaxation rate is given by
$t_1,t'_1 \sim t$ and that the negative frequency part of the
correlator $\langle g_i (t+\tau) g_i^*(t)\rangle $ comes from the negative frequency
term $A_t^2 \exp(i\omega_l (t_1 - t'_1))/4$ of the $Q(t_1)Q(t_{1'})$ (
here $t_1$ and $t'_1$ are the integral variables in (7) and (8),
$t_1 > t'_1$).

\section{Two-phonon decay}
In the case of two-phonon decay ($n=2$) the equations
(7) are linear. These equations have been solved in \cite{hizh1} which allowed us
to derive the explicit expression for the relaxation rate.
Here we present another
derivation of the formula for the relaxation rate
which is based on the direct calculation of the pair correlation functions
$D^{(1)}(t;\omega) \equiv D(t;\omega)$ and which can be extended for
three- and more- phonon decay processes.

Taking into account (8) and (11) one gets the following equation
for $D(t;\tau)$:
\begin{eqnarray}
D(t;\tau) &=& d^* (t;-\tau) + 2A_t \int_0^{t_0}dt_1
G(t_0 -t_1) V^{(3)} D(t; t_1 - t) \cos{\omega_l t_1},
\end{eqnarray}
where $t_0 = t+\tau, d(t;\tau) = \langle g(t+\tau) f^*(t)\rangle$.
For $t \gg \tau \lesssim \omega_l^{-1}$ the matrix-function $d(t;\tau)$
satisfies the same equation as $D(t;\tau)$ with
$d_0(\tau) = \langle f(\tau) f^*(0)\rangle $ instead of $d^* (t;-\tau)$. This equation
(as well as (12)) can be solved as follows. First we present this equation in
the form
\begin{eqnarray}
d(t;\tau) &=& d_0(\tau) + 2A_t \int_0^{t_0}dt_1 G(t_0 - t_1)V^{(3)} d_0
(t_1 - t)\cos{\omega_l t_1} + \nonumber \\
&& 4A_t^2\int_0^{t_0}dt_1\int_0^{t_1}dt_2
G(t_0 -t_1) V^{(3)}G(t_1 - t_2)V^{(3)} d(t; t_2 - t)\cos{\omega_lt_1}
\cos{\omega_lt_2}.
\end{eqnarray}
We are interested only in the positive frequency parts of
$D(t;\tau)$ and $d(t;\tau)$. The linear (with respect to $A_t$) term in the
right-hand side of (13) gives only the fast oscillating contribution to these
parts and it averages out. Omitting also the fast oscillating part of the
last two terms in (13) one gets the following equation:
\begin{equation}
d(t;\tau) \simeq d_0(\tau) + A_t^2 \int_0^{t_0} dt_1 \int_0^{t_1}dt_2
G(t_0 - t_1) V^{(3)} G(t_1 - t_2) V^{(3)} d(t;t_2-t)e^{i\omega_l(t_2-t_1)}
\end{equation}
Introducing
\begin{equation}
d_{(0)}(t;\omega) = \int_{-t}^{\infty}e^{i\omega \tau}
d_{(0)}(t;\tau)
\end{equation}
one obtains
\begin{eqnarray}
d(t;\omega) &=& d_0(t;\omega) + A_t^2 \int_0^{\infty}dt_0 \int_0^{t_0}dt_1
\int_0^{t_1}dt_2 e^{i\omega (t_0-t)} \nonumber \\
&&G(t_0 - t_1 )V^{(3)} G(t_1 - t_2 )
V^{(3)} d(t; t_2 - t ) \exp(i\omega_l(t_2 - t_1))
\end{eqnarray}
Denoting $\bar{t}_0 = t_0 - t_1$, $\bar{t}_1 = t_1 - t_2$ and taking into
account that $$
\int_0^{\infty}dt_0\int_0^{t_0}dt_1\int_0^{t_1}dt_2 =
\int_0^{\infty}dt_2 \int_0^{\infty}d\bar{t}_1
\int_0^{\infty}d\bar{t}_0 $$
one gets simple algebraic equation for $d(t;\omega)$ with the
following solution:
\begin{equation}
d(t; \omega) = [I - A_t^2 G(-\omega) V^{(3)} G(\omega_l - \omega) V^{(3)}]^{-1}
d_0 (t; \omega)
\end{equation}
Equation (12) can be solved in the same way. As a result one finds
\begin{equation}
D^{(1)}(t;\omega) = \frac{2}{\pi}
|I - A^2_t G(-\omega ) V^{(3)} G(\omega_l - \omega) V^{(3)}|^{-2}
ImG(\omega)
\end{equation}
and
\begin{eqnarray}
\frac{dE_{ph}(t)}{dt} &=& -\frac{dE_{l}(t)}{dt} \simeq
\frac{\hbar \omega_l A^2_t}{2\pi} \int_0^{\omega_M} d\omega \mbox{Sp}
[V^{(3)} Im G(\omega_l - \omega)  \nonumber\\
&&V^{(3)} |I-A_t^2 G(\omega-\omega_l)V^{(3)}
G(\omega) V^{(3)}|^{-2} ImG(\omega)].
\end{eqnarray}
Within notations and cyclic permutation of the matrices under Sp this
formula coincides with analogous formula derived in \cite{hizh1}
by another method. Note that diagonal elements of the resolvent-matrix
$[I-A_t^2 G(\omega - \omega_l) V^{(3)} G(\omega) V^{(3)}]^{-1}$ are
symmetric with respect to the change
$\omega \leftrightarrow \omega - \omega_l$.
Taking also into account that $G(-\omega) = G^*(\omega)$ one finds
that the resolvent
$([I-A_t^2G(-\omega_l/2)V^{(3)}G(\omega_l/2)V^{(3)}]^{-1})_{mm}$
is real. Therefore, the function under integral (19) for some critical values of
$A_t$ diverges at $\omega = \omega_l/2$. This leads to sharp peaks of the
relaxation
rate near $A_{cr}$ of the shape $|A_t - A_{cr}|^{-1} \sim |t-t_{cr}|^{-1/2}$
where  time moment $t_{cr}$ of the jump corresponds to $A_{t} = A_{cr}$
 \cite{hizh1}. The relaxation jumps are accompanied by explosion-like
emission of phonon bursts.

\section{Three- and more-phonon decay}
Let us consider now decay processes with creation of more than two phonons,
caused by quartic and higher order anharmonicities.
Substituting (8) to (11) one obtains after averaging over
$\phi_i$ the following nonlinear integral equations for
the matrix-function $D(t;\tau)$:
\begin{equation}
D(t; \tau) = d_0 (\tau) + 4(n-1)\int_0^{t_0} dt_1 \int_0^{t} dt_2
Q(t_1) Q(t_2) G(t_0-t_1) V^{(n)} D^{(n-1)}(t;t_1 - t_2) V^{(n)} G(t-t_2) ,
\end{equation}
where $D^{(n-1)}(t;\tau)$ is matrix with the elements
$(D^{(n-1)}(t;\tau))_{mm'} = D^{n-1)}_{mm'}(t;\tau)$
(in (20) it is accounted that terms $\sim
\langle f(t+\tau)(g^{n-1}(t'))^*\rangle $ and
$\langle g^{n-1}(t')f^*(t)\rangle $ turn to zero for
$n \geq 3$ when averaging over random phases $\phi_i$). The Green functions
$G(\tau)$ essentially differ from zero only for small
$|\tau| \lesssim \omega_l^{-1}$. Therefore, in the limit
$t \gg \tau \lesssim \omega_l^{-1}$ one can replace the lower
limits of integration by $-\infty$.
Neglecting also fast oscillating terms $\sim \cos{(\omega_l(t_1+t_2))}$
one gets
\begin{eqnarray}
D(t;\tau) &\simeq& d_0 (\tau) + 2(n-1)A_t^2 \int_{0}^{\infty}d\tau_1
\int_{-\tau_1}^{\infty}dx \cos{(\omega_l(\tau - x))} \nonumber\\
&&G(\tau_1 + x) V^{(n)} D^{(n-1)}(t;\tau -x) V^{(n)}G(\tau_1)
\end{eqnarray}
(here $\tau_1 = t- t_2$, $x=t_0-t_1-\tau_1$). Taking into account
that
\begin{equation}
\int_0^{\infty}d\tau_1 \int_{-\tau_1}^{\infty}dx =
\int_0^{\infty}dx \int_0^{\infty} d\tau_1 +
\int_{-\infty}^{0}dx
\int_{-x}^{\infty}d\tau_1 \nonumber
\end{equation}
and introducing the fourth-rank matrix-function
\begin{equation}
\tilde{G}_{mm_1;m'm_2} (x) = \int_0^{\infty} d\tau_1 G_{mm_1}(\tau_1+x)
G_{m'm_2} (\tau_1)
\end{equation}
one gets the following set of nonlinear integral equations:
\begin{eqnarray}
D_{mm'}(t; \tau ) &=& d_{0,mm'}(\tau ) +\nonumber \\
&&2(n-1)A_t^2 \sum_{m_1 m_2} V_m^{(n)} V_{m'}^{(n)} \int_{-\infty}^{\infty}
dx \cos{(\omega_l x)} D_{m_1 m_2}^{n-1}(t;x) \tilde{G}_{mm_1;m'm_2}(|\tau -x|)
\end{eqnarray}
($n \geq 3$). These equations can be solved
numerically and, in some cases, analytically. Then, performing Fourier
transformation (10) one can find $D^{(n-1)}(t;\omega)$ and then the
relaxation rate (9).

\section{Switch-off of multiphonon processes}
As an example, let us consider a $n$-phonon decay ($n \geq 3$) of a high
frequency local vibration, directed to the
nearest neighbours. In this case only two coordinates $q_m$ of the
nearest neighbours
can be accounted. Besides, the high frequency local mode breaks the dynamical
correlation (nonorthogonality) of these coordinates. Due to that the
Green functions $G(\tau)$ and $\tilde{G} (|x|)$ become diagonal. Therefore
the equation (24) takes the form
\begin{equation}
D(t; \tau) = d_{0}(\tau) + 2(n-1)V^{(n)^{2}}A^2_t
\int_{-\infty}^{\infty}dx D^{n-1}(x)\tilde{G}(|\tau - x|)\cos{\omega_lx}
\end{equation}
(index $m$ is omitted). We consider the case of phonon spectrum with narrow
single maximum at $\omega_0 = \omega_l/n$ of the Lorentzian shape
with the
width $\Gamma (\ll \omega_0)$.

In this case
\begin{equation}
G(\tau) = \omega_0^{-1}e^{-\Gamma |\tau|}\sin{\omega_0\tau};
d_0(\tau) = \omega_0^{-1} e^{-i\omega_0\tau - \Gamma |\tau |};
\tilde{G}(x) = \frac{1}{4\Gamma\omega_0^2}e^{-\Gamma|x|}\cos{(\omega_0 x)}
\end{equation}
Approximate solutions of the integral equation (25) is
\begin{equation}
D(t;\tau) \approx \omega_0^{-1} [\alpha_t e^{-i\omega_0\tau} +
\beta_t e^{i\omega_0 \tau}]e^{-\Gamma |\tau|}
\end{equation}
($t \gg \tau \sim \omega_0^{-1}$), where
$\beta_t = w_t\alpha_t^{n-1}$,
\begin{equation}
w_t = \frac{\hbar^{n-2}V^{(n)^{2}} A_t^2 (n-1)^2}
{2\omega_0^{n} \Gamma^2 n(n-2)};\;\;\;n\geq 3
\end{equation}
$\alpha_t$ is a solution of the equation
\begin{equation}
\alpha_t = 1 + w_t^n \alpha_t^{2(n-1)} .
\end{equation}
\vspace{3mm}

\begin{center}
{\bf TABLE 1}

\vspace{2mm}

\begin{tabular}{|c||c|c|c|c|c|c|c|c|} \hline
{\bf n}&{\bf 3}&{\bf 4}&{\bf 5}&{\bf 6}&{\bf 7}&{\bf 8}&{\bf 9}&{\bf 10}\\ \hline
$w_{n,cr}$ & 0.32 & 0.405 & 0.47 & 0.52&0.56&0.595&0.62&0.64\\ \hline
$\alpha_{n,cr}$&1.25&1.173&1.127&1.09&1.068&1.059&1.042&1.031\\ \hline
\end{tabular}

\vspace{3mm}

\end{center}This equation has real solutions only for $w_t$ being below the critical
value $w_{n,cr}$ and, correspondingly for $A_t < A_{n,cr}$ and
$\alpha_t < \alpha_{n,cr}$; values of $w_{n,cr}$ and $\alpha_{n,cr}$,
see in Table 1).

There are two such solutions but only the smaller
one corresponds to slow relaxation (as compared to $\omega_l$) and, therefore
only this solution should be accounted.
The relaxation rate, then equals
\begin{equation}
-\frac{dE_l}{dt} = \frac{A_t^2V^{(n)^{2}} \alpha_t^{n-1}}
{(\hbar\omega_0)^{n-1}\Gamma} = \hbar\omega_l\Gamma\alpha_t^{n-1}
w_t {\Big \lbrack} \frac{n-2}{n-1}{\Big \rbrack} ^2 .
\end{equation}
For small amplitude $A_t$, $\gamma = d \ln{E_{ph}(t)dt}$
does not depend on $A_t$, i.e. relaxation is exponential in accordance
with perturbation theory \cite{klemens}.
For $A_t \leq A_{n,cr}$ the relaxation rate grows
as $A_t \rightarrow A_{n,cr}$
and gets rather high value $\sim \Gamma\hbar\omega_l$ (of the order of
the mode quantum per period).
Abscence of real solutions of the
equation (29) for $w_t > w_{n,cr}$ ($A_t > A_{n,cr}^{(n)}$) means that $n$-phonon relaxation
is fully switched-off. Note that in the case of two-phonon decay the
switch-off effect also exists but it takes place only asymptotically
with $A_t \rightarrow \infty$ \cite{hizh1}. The values of $A_{n,cr} \sim
\hbar^{1-n/2} w_{n,cr}^{1/2}\Gamma\omega^{n/2}/|V^{(n)}|$ for
$n \geq 3$ are much larger than the corresponding values of the critical
amplitudes for two-phonon decay (for $n \geq 3$ $ A_{n,cr} \rightarrow \infty$
if $\hbar \rightarrow 0$). Typically $A_{3,cr} \sim 1 \AA \ll A_{4,cr} \ll
A_{5,cr}$ etc. Note also that the the difference of the
perturbative and
nonperturbative treatments on the n-phonon decay for $n \geq 3$ is given by
the factor $\alpha_t^{n-1}$ in (30). For all $A_t < A_{n,cr}$ this factor
does not exceed 1.62. Therefore, the main results of the nonperturbative
consideration  of the multiphonon decay are the switching-off of the process
if the mode amplitude exceeds the corresponding critical value, and the
step-vise switching-on of the very fast relaxation when the amplitude
reaches this value from above. For not very large amplitudes ($< 1\AA$) three-
and more- phonon decay processes are well described by the
standard perturbation theory.

Thus, the general scenario of the relaxation of strongly excited local
mode is the following. In the beginning the relaxation (which takes place
due to creation of $n \geq 4$ phonons) is rather slow. Then,
when the mode amplitude approaches the critical value for the $(n-1)$-phonon
decay processes, relaxation is jump-like enhanced (supposing these processes
are allowed by the energy conservation law). After that the $(n-1)$-
phonon processes will slow-down till the next critical amplitude
(which coresponds to the $(n-2)$-phonon processes), then the
relaxation process will be once more jump-like enhanced (supposing the
$(n-2)$-phonon process is allowed) and so on. The relaxation rate
can change very fast not only for a small change of the mode amplitude
but also for a small change of the mode frequency.
Final stage of the anharmonic relaxation is
exponential and is well described by the perturbation theory.

In conclusion, we developed here a nonperturbative theory of multiphonon
anharmonic decay of a strong local mode. We found that three- and more-
phonon decay processes of the mode are described by essentially
different equations than
the two-phonon decay process. The main effect which is predicted by the
nonperturbative
theory is the switching-off of the $n \geq 3 -$phonon process
if the mode amplitude exceeds the corresponding critical value and sharp
(step-wise) switching-on of the fast relaxation when the amplitude
reaches this value from above.

The author is grateful to A.Bussman-Holder
for invitation to participate in the conference.
The research was supported by Estonian SF Grant No. 2274.

\end{document}